\documentclass[aps,prd,amssymb,showpacs,floatfix, bibliography, twocolumn]{revtex4-1}
\usepackage{amssymb,graphicx}
\usepackage{epsfig,graphicx}
\usepackage{subcaption}
\usepackage{subfloat}
\usepackage{epstopdf}
\usepackage{amsmath}
\usepackage{tikz}

\begin{document}

\title{Probing topological properties of 3D lattice dimer model with neural networks}

\author{Grigory Bednik}
\affiliation{ UC Santa Cruz Physics Department, 1156 High Street Santa Cruz, CA 95064, USA}


\begin{center}
\begin{abstract}

Machine learning methods are being actively considered as a new tool of describing many body physics. However, so far, their capabilities has  been only demonstrated  in previously studied models, such as e.g. Ising model.
Here, we
consider a simple 
 problem, demonstrating 
that neural networks can be successfully used to give new insights 
in statistical physics.
 Specifically, we consider 3D lattice dimer model, which consists of sites forming a lattice and bonds connecting the neighboring sites, in such a way that every bond can be either empty or filled with a dimer, and the total number of dimers ending at one site is fixed to be one.  
Dimer configurations can be viewed as equivalent if they are connected through a series of local flips, i.e. simultaneous 'rotation' of a pair of parallel neighboring dimers. It turns out that the whole set of dimer configurations on a given 3D lattice can be split into distinct topological classes, 
such that dimer configurations belonging to different classes are not equivalent.  In this paper we identify these classes by using neural networks. More specifically, we train the neural networks to distinguish dimer configurations from two known topological classes, and after it, we test them on dimer configurations from unknown topological classes. We 
 demonstrate that 3D lattice dimer model on a bipartite lattice can be described by an integer topological invariant ('Hopf number'), whereas lattice dimer model on a non-bipartite lattice is described by $Z_2$ invariant.  
Thus, we demonstrate that neural networks can be successfully used  to identify new topological phases in condensed matter systems, whose existence can be later verified by other (e.g. analytical) techniques. 

\end{abstract}
\end{center}

\maketitle
\section{Introduction}

A key feature of physics, which distinguishes it from other areas of human knowledge is that it describes the world in terms of exact physical laws. Dynamics of any system can, in principle, be reduced to dynamics of its components and their interactions, and each physical component can be exactly described in terms of its equations of motion. However, this simple paradigm can be difficult to realize in practice, especially when the number of the consitutent components becomes too large. Statisical physics deals with systems containing close-to-infinity number of elements, which makes it somewhat similar to various statistical problems, 
where individual laws are not known, but the only known thing 
is a small sample of the ensemble and its empirical properties. These properties may be, in principle, uncovered by using statistical learning - a way to extract general patterns within the sample and to use them to predict the properties of the whole system. Statistical learning includes various methods \cite{james2014introduction}, and the most prominent of them are neural networks \cite{Goodfellow-et-al-2016}. From this perspective, it seems interesting to use them to get knowledge about statistical physics. Indeed, in the recent years, neural networks were proven to 
be successful in describing
various physical properties, such as magnetic phases and critical temperature of the 
 Ising model \cite{Carrasquilla2017}, as well as other similar lattice models (such as e.g. Hubbard model \cite{PhysRevX.7.031038}, topological phases in various models \cite{PhysRevLett.120.066401, zhang2017machine, beach2018machine}, many body localization \cite{PhysRevLett.120.257204} etc). We also remark that neural networks have also been used in condensed matter physics for other various purposes (e.g. representaion of quantum states \cite{PhysRevB.96.195145} or many-body quantum state tomography \cite{Torlai2018}). However, the key question is: can neural networks give new insights about many body physics, which were not discovered before?

To address this question, we consider another well-known model of statistical physics - 3D lattice dimer model, which contains sites forming a lattice and edges connecting the sites, so that every edge can be either empty or occupied by a dimer, provided the total number of dimers at each site is equal to one. This simple model has a long history in condensed matter physics. It has been studied since 1960-s, when it was realized that the total number of dimer coverings on a planar lattice can be computed analytically \cite{doi:10.1063/1.1703953}. Soon after, it was realized that dimer model on a so-called Fisher lattice is dual to 2D Ising model on a square lattice, which made it possible to solve the Ising model analytically \cite{doi:10.1063/1.1704825}. Later on, lattice dimer model has been extensibely probed as a candidate model for high-temperature superconductivity \cite{ANDERSON1196, PhysRevLett.61.2376} - it was suggested that dimers may describe electron singlets formed between neighboring sites. More recently, dimer model on a diamond lattice has been interpreted as a dual to spin system on pyrochlore lattice, which, in turn, hosts exotic spin ice state \cite{PhysRevB.84.115129, PhysRevB.69.064404, doi:10.1146/annurev-conmatphys-020911-125058, RevModPhys.82.53}. The most up-do-date idea to realize lattice dimer model in experiments is a so-called artificial spin ice, i.e. a lattice of nanomagnets \cite{RevModPhys.85.1473, Perrin2016, Lao2018, doi:10.1063/1.4861118, Keller2018}.

Dynamics of the lattice dimer model can be described by applying local flips to dimers, i.e. change of orientation for a pair of parallel dimers within one plaquette (see Fig. \ref{DimerFlip}). This definition of dynamics is natural given that all condensed matter systems are local. Moreover, the concept of local flips follows from quantum generalization of the lattice dimer model, known as Rokhsar-Kivelson model \cite{PhysRevLett.61.2376}. The later is known to host a spin liquid phase \cite{PhysRevB.68.184512}, whose excitations spectrum depends on the type of the underlying lattice: on a bipartite lattice gapless $U(1)$ phase is realized, whereas on a non-bipartite lattice, a gapped $Z_2$ phase appears. The difference between bipartite and non-biparite lattices persists on a classical level: on a bipartite lattice, dimer configurations can be described using effective magnetic field (which in 2D case gets reduced to 'weights representation') \cite{PhysRevLett.91.167004}, whereas a similar  representation is not known on a non-bipartite lattice.

Despite its simplifity, theoretical properties of the classical lattice dimer model are not yet fully explored. For example, one important question about it 
 is: can we use a sequence of local flips in order to transform one given configuration of dimers to another given configuration? This question was in part addressed in the Ref. \cite{PhysRevB.84.245119} - it was pointed out that configurations in lattice dimer model can be characterized by an invariant, which does not change under local flips, a pfaffian of so-called Kasteleyn matrix. It was found that in the case of a planar lattice (e.g. 2D plane), such invariant is always equal to $+1$, whereas in the case of a non-planar (e.g. 3D) lattice, there exist both 
configurations,  
 where this invariant is 
 equal to 
$+1$ (such as trivial maximally flippable state, see Fig. \ref{Hopfion_0}),  
 and 
$-1$ (a 'hopfion'  - the name was introduced from a continuum limit of a cubic lattice \cite{PhysRevB.84.245119}, see Fig. \ref{JustHopfion}).

The problem of finding distinct
topological
 classes of dimer configurations 
can be viewed as a classification problem from machine learning point of view,
and therefore, it 
can be addressed by using the most powerful machine learning method - neural networks. 
Motivated by this,
we study dimer configurations in the following way: 
first, we train the neural network on a dataset of configurations from two known topological classes, 
and after it, we test the neural network on a different dataset of 
 configurations from various topological classes.
We find that the neural network is able to successfully distinguish dimer configurations from the two topological classes used for training, 
but more interestingly, the neural network is able to distinguish 
dimer configurations from other classes, 
thus answering the question of the full topological classification.

 We obtain that 
on a bipartite (in our case cubic) lattice, hopfions are characterized by an integer topological invarariant, whereas on a non-bipartite lattice (we consider an example of stacked triangular lattice), dimer configurations are characterized by $\mathtt{Z}_2$ invariant. We remark, that this reasoning gave us a hint that on a bipartite lattice, dimer configurations can be characterized by an exact Hopf number \cite{e1996force, arnold2013topological, PhysRevB.84.184501, PhysRevLett.51.2250, PhysRevB.88.201105, PhysRevB.94.035137, 0256-307X-35-1-013701, PhysRevB.95.161116, PhysRevLett.119.156401, Volovik1977}, which 
we later verified analytically  \cite{MyArticle}.
However, on a non-bipartite lattice, the neural network is the only known way to obtain the topological classification of dimer configurations. 

This is the main
idea of the paper:
 neural networks can  
successfully
identify  new topological phases, not known in advance, and as such, they can be
used to give 'hints' about unknown properties of physical systems,
which can be later
verified
by other, more rigorous 
techniques.  

This paper is organized as follows. In Sec. \ref{Sec:CubicLattice} we introduce 3D lattice dimer model and describe our method in the case of cubic lattice. In Sec. \ref{Sec:TriangularLattice} we repeat our study in the case of stacked triangular lattice. In Sec. \ref{Sec:Discussion} we summarize our findings. In the appendix, we describe technical details of our method (Sec. \ref{Sec:Methods}) and briefly present analytical construction of the Hopf invariant (Sec. \ref{Sec:HopfNumber}). For a more detailed discussion about Hopf invariant in 3D lattice dimer model, we refer to the Ref. \cite{MyArticle}.


\section{Dimer model on a cubic lattice}
\label{Sec:CubicLattice}

\begin{figure}
	\begin{subfigure}[b]{0.2\textwidth}
		\includegraphics[width=4cm, angle=0] 
		{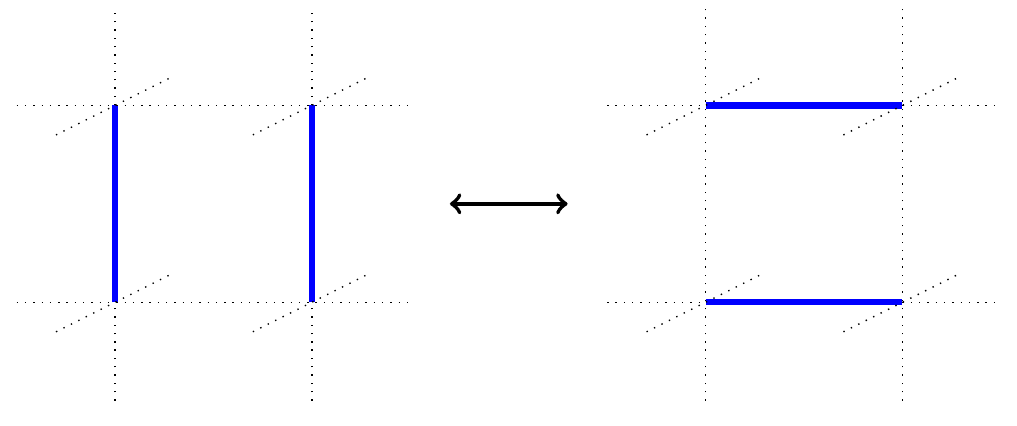}
		\subcaption{}	
		\label{DimerFlip}
	\end{subfigure}
	\hspace{1cm}
	\begin{subfigure}[b]{0.2\textwidth}
		\includegraphics[width=4cm,angle=0] 
		{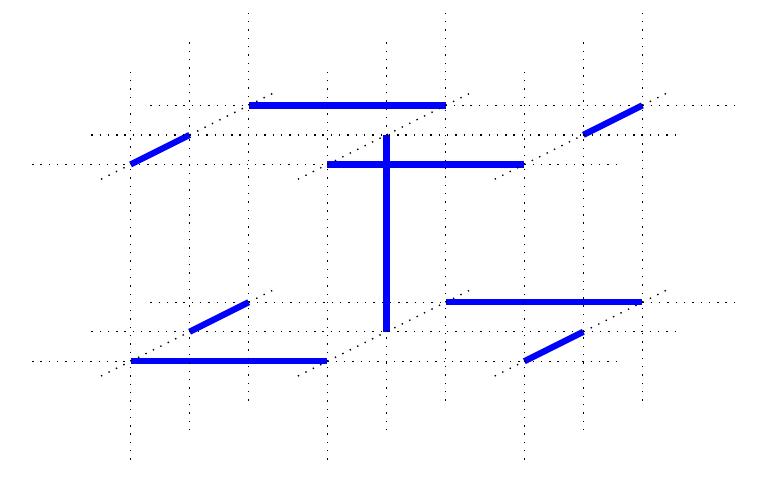}
		\subcaption{}	
		\label{JustHopfion}
	\end{subfigure}
	\hspace{1cm}
	\begin{subfigure}[b]{0.2\textwidth}
		\includegraphics[width=4cm,angle=0] 
		{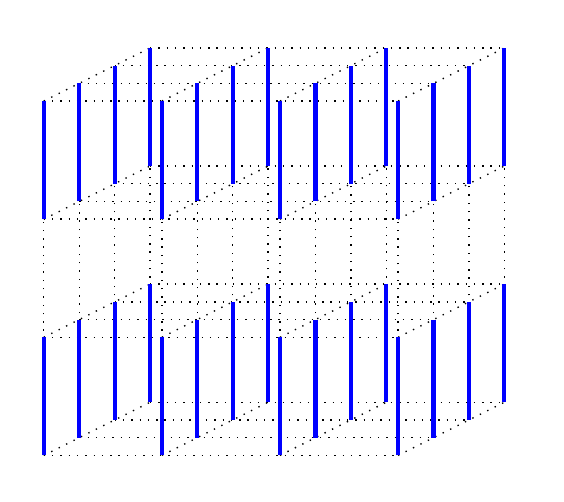}
		\subcaption{}	
		\label{Hopfion_0}
	\end{subfigure}
	\hspace{1cm}
	\begin{subfigure}[b]{0.2\textwidth}
		\includegraphics[width=4cm,angle=0] 
		{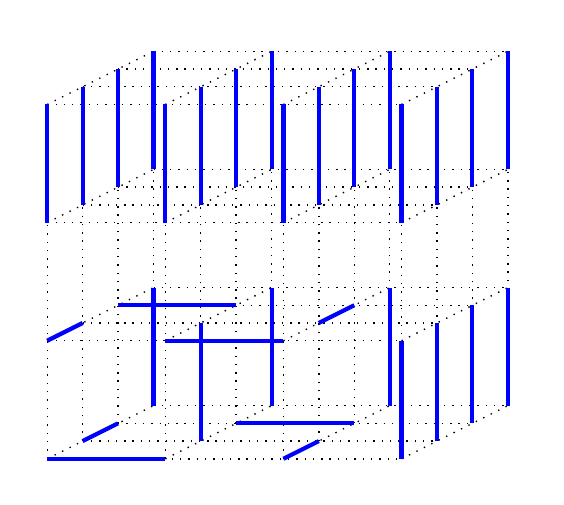}
		\subcaption{}	
		\label{Hopfion_1}
	\end{subfigure}
	\hspace{1cm}
	\begin{subfigure}[b]{0.2\textwidth}
		\vspace{-0.4cm}
		\includegraphics[width=4cm,angle=0] 
		{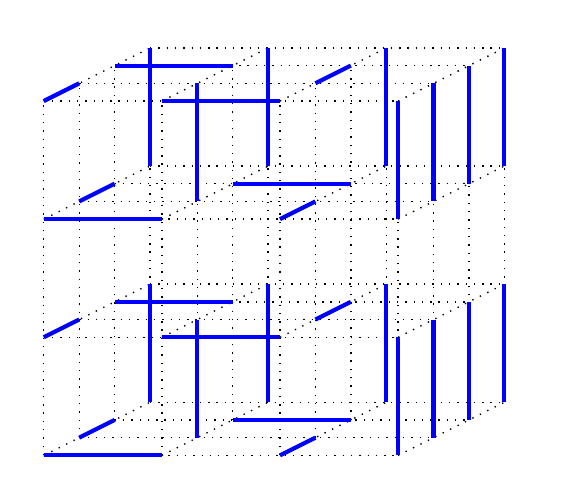}
		\subcaption{}	
		\label{Hopfion_2}
	\end{subfigure}

\caption{
A lattice dimer model consists 
of a dimer lattice lattice, and dimers 
 placed on its bonds, in such a way that every site is attached to exactly one dimer. (\subref{DimerFlip}) Two dimer configurations are considered equivalent if they can be connected to each other by a series of local flips.  (\subref{JustHopfion}) The simplest dimer configuration, which is not equivalent to trivially aligned dimers is a hopfion. To perform our study we used  (\subref{Hopfion_0}) $4 \times 4 \times 4$  lattice with trivially aligned dimers, (\subref{Hopfion_1}) a hopfion placed on a lattice of the same size and surrounded by trivially aligned dimers, and (\subref{Hopfion_2}) two hopfions stacked on the same lattice, and also surrounded by trivially aligned dimers. 
}
\label{DimerConfigurations}
\end{figure}

We start from revising the basic properties of lattice dimer model. Let us consider a lattice, i.e. a chain of periodically aligned sites, and assume that each pair of the nearest neighboring sites is connected with a bond. We place dimers on some of the bonds, i.e. assume that every bond is either empty, or occupied with a dimer. We also assume that the dimers are placed on a lattice in such a way, that they satisfy a constraint: every site is attached to exactly one dimer. 
A  few possible examples of dimer configurations on a lattice with $4 \times 4 \times 4$ sites are shown on the Fig. \ref{DimerConfigurations}. 
Indeed, on the Fig. \ref{DimerConfigurations} one can see the bonds filled with dimers, and check that every lattice site is attached to exactly one dimer.

We allow dimer configurations to change by applying random local flips. 
 A local flip is a transformation, which simultaneously changes orientation of two parallel dimers within one plaquette (see Fig. \ref{DimerFlip}). We refer to a pair of  configurations as equivalent, if they can be trasnsformed into each other by a series of local flips. For example, on the Fig. \ref{Hopfion_0} all dimers are parallel to each other, and therefore it is possible to apply a local flip to any of its plaquettes. Afterwards, one can repeat applying local flips to 
to any of the plaquettes, whose dimers are parallel, thus generating various equivalent configurations. On the Figs. \ref{Hopfion_1}, \ref{Hopfion_2} not all dimers are parallel to each other, and therefore 
one can apply local flips to those plaquettes, whose dimers are parallel. 
Thus 
from every dimer configuration, it is possible to generate a lot of equivalent configurations by applying local flips. 

\begin{figure}
	\begin{subfigure}[b]{0.2\textwidth}
		\includegraphics[width=4cm, angle=0] 
		{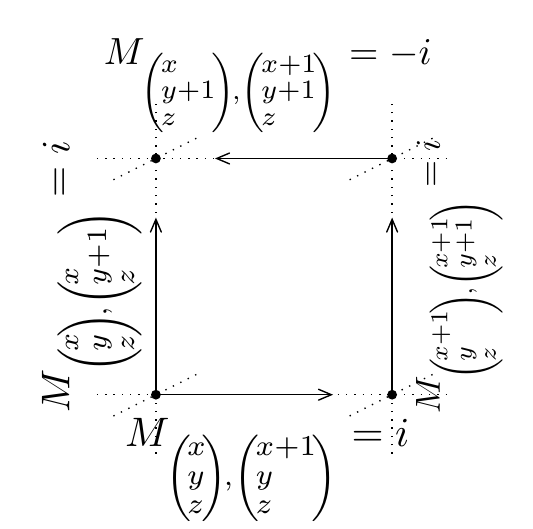}
		\subcaption{}	
		\label{MatrixArrangementPlaquette}
	\end{subfigure}
	\hspace{1cm}
	\begin{subfigure}[b]{0.2\textwidth}
		\vspace{0.4cm}
		\includegraphics[width=4cm,angle=0] 
		{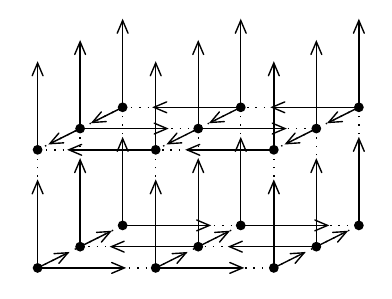}
		\subcaption{}	
		\label{MatrixArrangement}
	\end{subfigure}
	\hspace{1cm}
\caption{\small  A dimer lattice is parametrized by antisymmetric matrix $M_{ij}$, whose components are equal to $\pm i$ in such a way, that a product of matrix components around a plaquette is equal to $-1$.
(\subref{MatrixArrangementPlaquette}) shows arrangement of signs of $i$ along a given plaquette, which are marked by arrows. (\subref{MatrixArrangement}) shows the same arrangement within 3D lattice.
\label{KasteleynMatrix}
}
\end{figure}

The key question, that we want to answer is: are all dimer configurations on a given lattice are equivalent, i.e. can be obtained from each other by applying local flips, or are there distinct topological classes, such that configurations from different classes cannot be transformed into each other by applying local flips? Previously, in the Ref.  \cite{PhysRevB.84.245119}, there was presented an argument that not all configurations in 3D lattice dimer model are equivalent. The idea was the following: the lattice can be characterized by Kasteleyn matrix $M_{ij}$ with the indices $i,j$ enumerating all lattice sites, such that its components take values $\pm i$, and 
their signs are chosen in such a way, that a product of $M_{ij}$
 around each plaquette is equal to $-1$ 
(see Figs. \ref{MatrixArrangementPlaquette}, \ref{MatrixArrangement} for the precise arrangement).
Similarly, each dimer configuration can be characterized by another matrix $n_{ij}$, whose components are equal to $1$ if the sites $i,j$ are connected with a dimer, and zero otherwise. It is straightforward to check that the expression $\mathrm{Pf}  (M_{ij} n_{ij})$ is invariant under local flips. On the other hand, one can explicitly compute this invariant for specific dimer configurations and see that it may take different values. For example, this invariant is equal to $1$ for a trivial dimer configuration, shown on the Fig. (\ref{Hopfion_0}). In contrast, a non-trivial configuration, shown on the Figs. (\ref{JustHopfion}, \ref{Hopfion_1}) has $\mathrm{Pf}  (M_{ij} n_{ij}) = -1$. 
Since the invariant $\mathrm{Pf}  (M_{ij} n_{ij})$ does not change under local flips, but at the same time it takes different values for two configurations  '0' (Fig.\ref{Hopfion_0}) and '1' (Fig.\ref{Hopfion_1}), these configurations cannot be transformed to each other by applying a series of local flips. We refer the configuration shown on the Fig. \ref{JustHopfion} as a \textit{hopfion}, following the Ref. \cite{PhysRevB.84.245119}, where it was given such name, because in the continuum limit, it behaves as a field configuration with a non-trivial Hopf number. 



The previous argument makes it possible to see that space of all dimer configurations contains different inequivalent classes, but 
there still remains a question whether a pair of dimer configurations with the same value of $\mathrm{Pf}  (M_{ij} n_{ij})$ always belong to the same class. For example, if we 'stack' two hopfions on top of each other (see Fig. \ref{Hopfion_2}), the resulting configuration has $\mathrm{Pf}  (M_{ij} n_{ij}) =+1$, i.e. the same as for the trivial dimer configuration (Fig. \ref{Hopfion_0}), but do they belong to the same topological class? To find an answer to this question, we implement one of the most popular machine learning algorithms - a neural network. The main idea is the following: if we train the neural network to distinguish configurations equivalent to the trivial (Fig. \ref{Hopfion_0}) and the hopfion (Fig. \ref{Hopfion_1}), what will it tell us about the unknown configuration containing two hopfions (Fig. \ref{Hopfion_2})?

We create our training and test datasets by Monte Carlo method.  
More specifically, we consider a  cubic  lattice of the size of $4 \times 4 \times 4$ and with open boundary conditions. We start from 
configurations from each of the classes, shown on the Figs. (\ref{Hopfion_0} - \ref{Hopfion_2}) and apply a sequence of local flips to each of these two configurations. In particular, to generate the training dataset, we start from two configurations '0' and '1'  from each of the classes: the first  
 ('0') has all dimers aligned in $z$ direction (see Fig. \ref{Hopfion_0}), and the second 
 ('1') is a hopfion surrounded by vertically aligned dimers (see Fig. \ref{Hopfion_1}). We apply a sequence of local flips to each of these two configurations, and assign the label '0' or '1' to the outputs by using the fact that local flips preserve the topological class.

We train the neural network to distinguish, whether each configuration belongs to the class '0' or '1'. More specifically, we define the neural network in such a way, that it takes a dimer configuration as input, and outputs its topological class. We use a fully-connected neural network with one hidden and output layer, and repeat the procedure for different number of hidden units. 
In each hidden unit, we use the activation function $relu$, but in the output layer,  
 we do not use any activation function: in other words, the output is simply a linear superposition of the results from hidden units with an added bias.

After training, we apply the neural network
 to a test dataset, which is generated  by applying local flips to  configurations containing zero, one and two hopfions respectively (see Figs. \ref{Hopfion_0}, \ref{Hopfion_1}, \ref{Hopfion_2}). We find, that our neural network can successfully distinguish all of them. It outputs a real number approximately equal to
the number of hopfions (which can be either zero, or one, or two), and its accuracy improves with increasing number of units. 
Thus, if we assume that the trivial configuration (Fig. \ref{Hopfion_0}) has topological number 0, and configuration with one hopfion (Fig. \ref{Hopfion_1}) has topological number 1, then the neural network tells us that the configuration with two hopfions (Fig. \ref{Hopfion_2}) has topological number 2. In other words, the neural network gives us a hint that the dimer configurations (Fig. \ref{Hopfion_0}, \ref{Hopfion_1}, \ref{Hopfion_2}) are characterized by an integer topological invariant, as we would expect from its continuum limit. Since, we know that in the continuum limit, field configurations are characterized by a Hopf invariant, we believe that our lattice configurations are characterized by the same integer topological  invariant.



\begin{figure}[h]
\centering
		\includegraphics[width=8cm,angle=0]{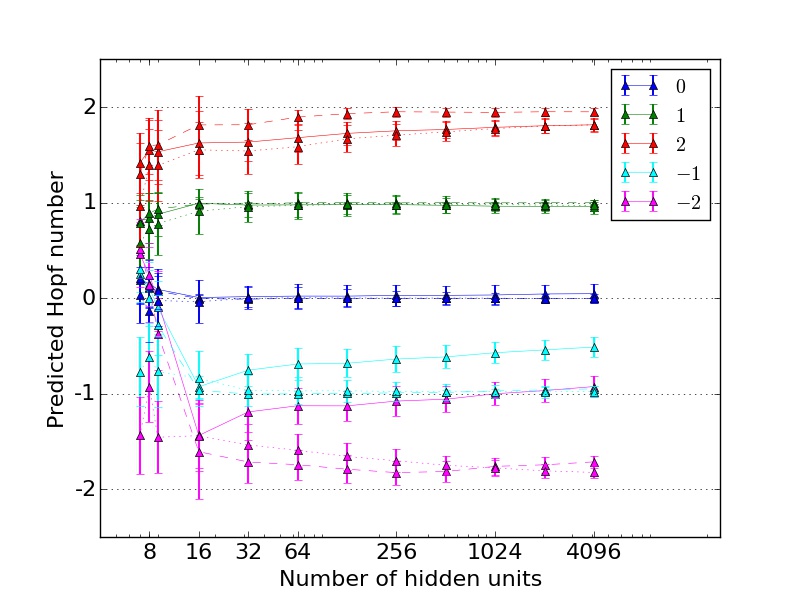}
\caption{\small Outputs of the neural network as functions of the number of hidden units applied to trivial configurations ("0"), generated from one hopfion ("1"), two of them ("2"), mirror reflected  hopfion ("-1"), mirror reflected pair of hopfions ("-2").  
 The case of solid line corresponds to the neural network trained on a  small dataset ($\sim 10^4$ training samples, 1500 epochs), containing configurations"0" and "1". Its accuracy on configurations with negative Hopf numbers can be improved by either increasing the training dataset (dashed line refers to the case of $\sim 6*10^6$ training samples, 145 epochs), or by retraining the neural network on configurations   "0" ,"1" and "-1", as shown by by the dotted line ($\sim 10^4$ samples, 200 epochs).
}
\label{HopfionPlots}
\end{figure}

If we believe, that a  hopfion is actually described by a  
 Hopf number, then its mirror image has to be described by Hopf number of the opposite sign. We are interested in checking it using our neural network. If we reflect the configurations within our test dataset, and substitute them into the neural network, it outputs negative number, which, for a small training dataset, does not as closely approach $-1$ or $-2$, as it approaches the positive integers describing hopfions without reflection. However, its accuracy increases with increasing number of samples in the training dataset. Thus we believe that at sufficiently large number of samples, the neural network trained on samples with Hopf numbers $0$ or $1$ should successfully recognize samples with Hopf numbers $2$, $-1$ and $-2$, though for the negative Hopf numbers, its accuracy improves slower over the size of the training sample, than for the positive Hopf numbers. We present our results on the Fig. \ref{HopfionPlots}. 

We note that recognizing samples with negative Hopf numbers can be improved by incorporating them into the training algorithm. Particularly, we can repeat our training procedure and to include three kinds of configurations in the training sample: trivial configurations with Hopf number $0$, configurations obtained from one hopfion with assigned Hopf number $+1$, and their mirror images with assigned Hopf number $-1$. In this case, the neural network can equally well recognize all configurations with Hopf numbers between $-2$ and $2$ (see Fig. \ref{HopfionPlots}). We believe that this result is consistent with an idea, that neural network can be successfully applied to samples within  the space, where it was trained, but, generally, it poorly extrapolates.


\subsection{Cubic lattice with periodic boundary conditions}

We are also interested, whether the fact that configurations in lattice dimer model can be characterized by Hopf number, depends on the kind of boundary condidtions imposed on the lattice. To find it out, we repeat our procedure in the case, when the lattice obeys periodic boundary condidtions, and we obtain similar results: if a neural network were trained on configurations with zero or one hopfion, it can successfully distinguish configurations obtained from zero, one, or two hopfions, but it has slightly lower accuracy when distinguishing their mirror reflected images, i.e. configurations with Hopf numbers $-1$ and $-2$. 
 The fact that Hopf number can be defined either on a lattice with open, or periodic boundary conditions is a non-trivial result, because a-priori, one might expect that topological properies of a model depend on topological properties of the manifold, where it is placed. Furthermore, we mention that in the work \cite{MyArticle} we develope a method of computing Hopf number analytically, but the idea presented there works only in the case of open boundary conditions. Thus, neural networks provide us with a qualitatively new result: dimer configurations on a lattice with periodic boundary conditions are characterized by Hopf number in the same way, as  
on a lattice with open boundary conditions. 

\begin{figure}[h]
\centering
		\includegraphics[width=8cm,angle=0] 
		{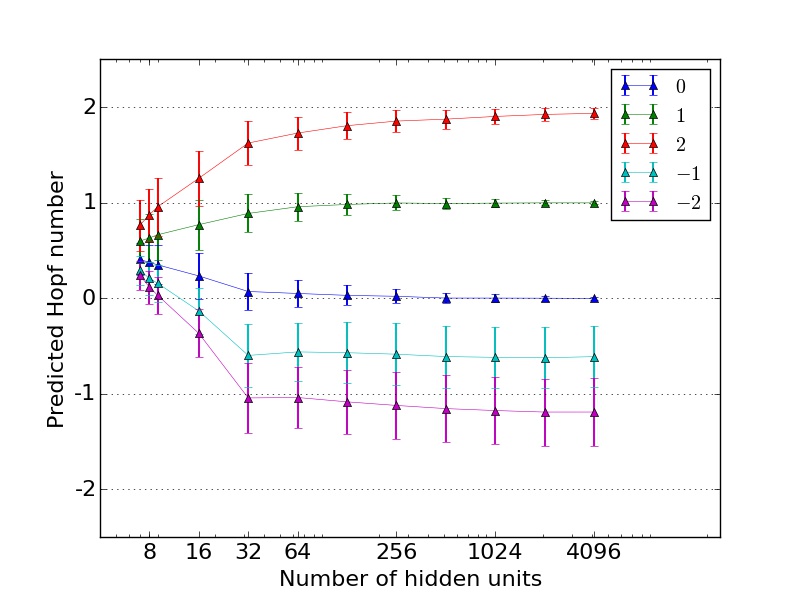}
\caption{\small Outputs of the neural network trained on configurations "0" and "1" for a $4 \times 4 \times 4$ lattice with periodic boundary conditions.   ($\sim 6*10^5$ samples, 500 epochs).
}
\label{Hopfion_Plot_Periodic}
\end{figure}

\section{Dimer model on a stacked triangular lattice}
\label{Sec:TriangularLattice}

\begin{figure}[h]
	\begin{subfigure}[t]{0.2\textwidth}
		\includegraphics[clip=true , trim = 1 1 1 1 , height=1.6cm  , width=3cm, angle=0]{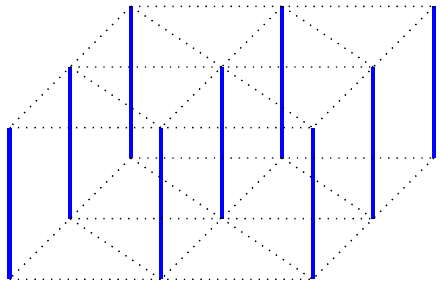}
		\subcaption{}
		\label{TrivialTriangular}
	\end{subfigure}
		\hspace{0.2cm}
	\begin{subfigure}[t]{0.2\textwidth}
		\includegraphics[clip=true , trim = 1 1 1 1 , height=1.6cm  , width=3cm,  angle=0]{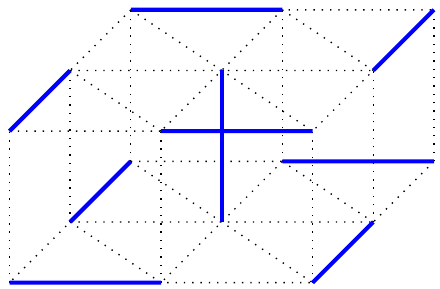}
		\subcaption{}	
		\label{JustHopfionTriangular}
	\end{subfigure}
	\begin{subfigure}[t]{0.2\textwidth}
		\includegraphics[clip=true , trim = 1 1 1 1 , height=1.2cm , width=3cm, angle=0]{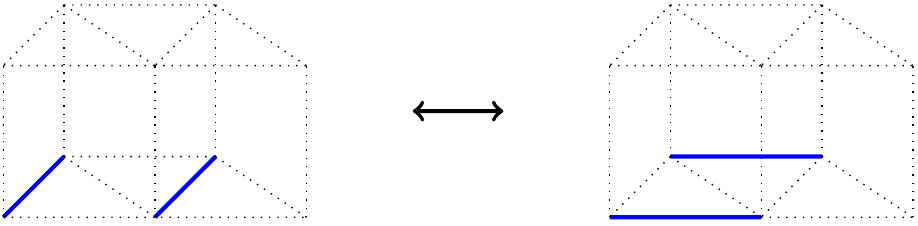}
		\subcaption{}
		\label{FlipTriangularOne}
	\end{subfigure}
		\hspace{1cm}
	\begin{subfigure}[t]{0.2\textwidth}
		\includegraphics[clip=true , trim = 1 1 1 1 , height=1.2cm  , width=3cm,  angle=0]{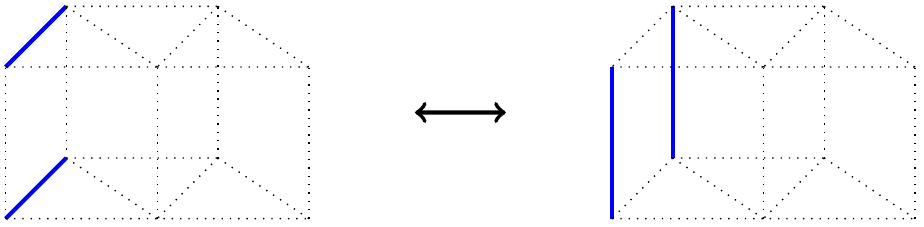}					
		\subcaption{}	
		\label{FlipTriangularTwo}
	\end{subfigure}
	\begin{subfigure}[t]{0.2\textwidth}
		\includegraphics[clip=true , trim = 1 1 1 1 , height=1.2cm  , width=3cm, angle=0]{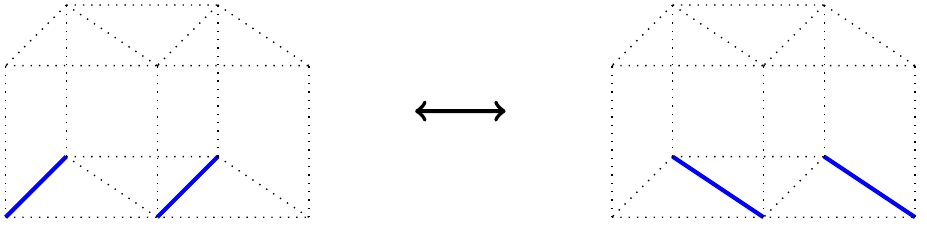}
		\subcaption{}
		\label{FlipTriangularThree}
	\end{subfigure}
		\hspace{1cm}
	\begin{subfigure}[t]{0.2\textwidth}
		\includegraphics[clip=true , trim = 1 1 1 1 , height=1.2cm  , width=3cm, angle=0]{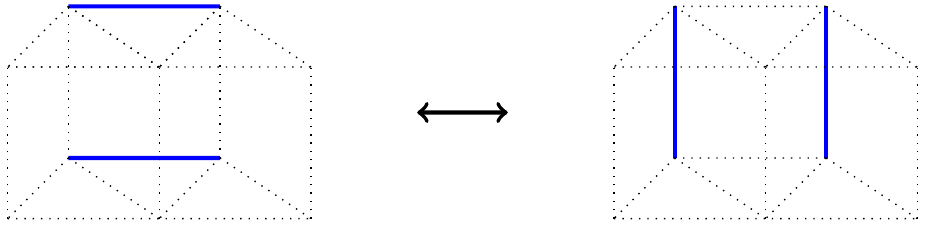}
		\subcaption{}	
		\label{FlipTriangularFour}
	\end{subfigure}
	\begin{subfigure}[t]{0.2\textwidth}
		\includegraphics[clip=true , trim = 1 1 1 1 , height=1.2cm   , width=3cm, angle=0]{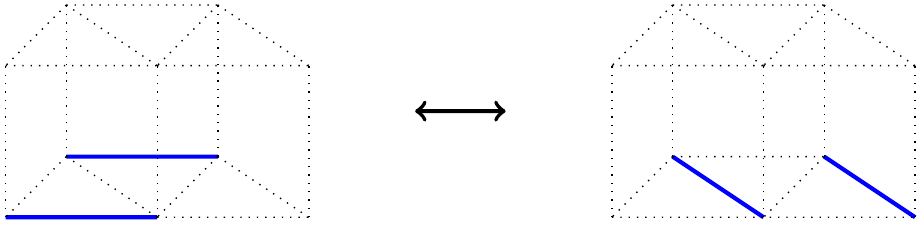}
		\subcaption{}
		\label{FlipTriangularFive}
	\end{subfigure}
		\hspace{1cm}
	\begin{subfigure}[t]{0.2\textwidth}
		\includegraphics[clip=true , trim = 1 1 1 1 , height=1.2cm   , width=3cm, angle=0]{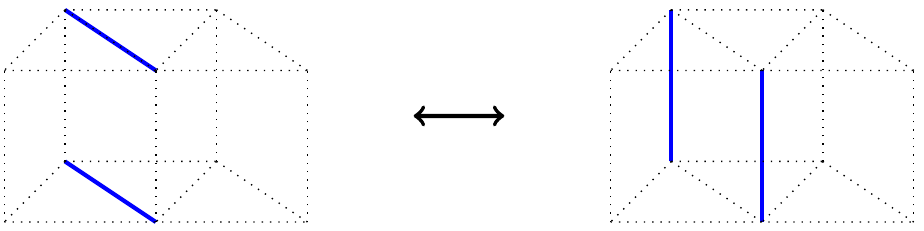}
		\subcaption{}	
		\label{FlipTriangularSix}
	\end{subfigure}
\caption{\small An example of stacked triangular lattice forming a (\subref{TrivialTriangular}) trivial dimer configuration and (\subref{JustHopfionTriangular}) a hopfion. (\subref{FlipTriangularOne}- \subref{FlipTriangularSix}) six possible local flips on a stacked triangular lattice that preserve $\mathrm{sign} \mathrm{Pf} (M_{ij} n_{ij})$.
}
\label{FlipTriangular}
\end{figure}

In the previous section we demonstrated that neural network can succesfully distinguish topological sectors of lattice dimer model on a cubic lattice. However, from previous studies of the lattice dimer model (e.g. \cite{PhysRevB.68.184512}), it is known that it has qualitatively different properties on a bipartite and non-bipartite lattices. For instance, the notion of effective magnetic field exists only if the lattice is bipartite, and furthermore, quantum dimer model has diffrent strongly coupled  
phases: on a bipartite lattice, it has a gapless $U(1)$ phase, but on a non-bipartite lattice a gapped $Z_2$ phase is realized. Motivated by this, we would like to apply our method to study hopfions on a non-bipartite lattice. We consider the most straightforward generalization of cubic lattice - stacked triangular lattice. It has the same sites and bonds as the cubic, but in addition, it has bonds aligned diagonally.  Thus, on a stacked triangular lattice, we can create initial configurations with zero, one or two hopfions in exactly the same way, as we did for a cubic lattice, but when we transform them, we apply more kinds of local flips: six kinds in total (see Fig. \ref{FlipTriangular}). 

As previously, we start from two initial configurations: the first with all dimers aligned in the vertical direction, and the second with one hopfion surrounded by vertically aligned dimers, and apply a series of local flips to both of them. In this way, we obtain a large number of configurations, which we use as a training dataset for the neural networks. 

\begin{figure}[h]
\centering
		\includegraphics[width=8cm,angle=0] 
		{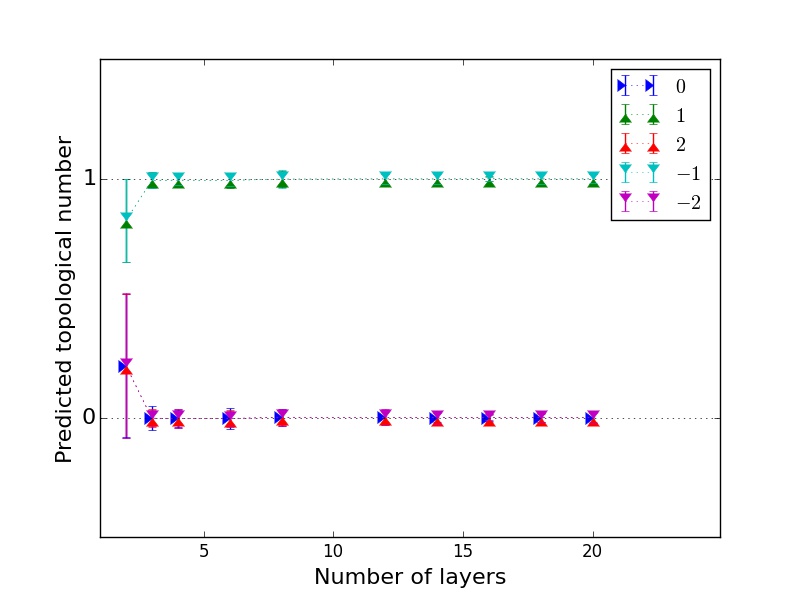}
\caption{\small Outputs of the neural network trained on configurations "0" and "1" for a $4 \times 4 \times 4$ stacked triangular lattice with open boundary conditions.   ($\sim 1.6*10^6$ samples, 400 epochs). One can see that the samples "0", "2", "-2" are indistinguishable. In the same way the samples "1" and "-1" are also indistinguishable. Thus we conclude that the neural network learns $Z_2$ invariant equal to the parity of the number of hopfions. 
}
\label{Hopfion_Plot_Triangular}
\end{figure}

After it, we create our test dataset by applying local flips to the dimer configurations '0' and '1' (trivial and a hopfion), as well as configurations with two hopfions and mirror reflected images of the configurations with one or two hopfions. We note, that in this setting, the neural network with only one hidden layer cannot be trained successfully, and therefore we have to increase the number of hidden layers, while keeping the number of units in each layer fixed. We obtain that the neural network with three or more layers can successfully distinguish configurations obtained from trivial dimer configuration or with configuration containing one hopfion (see Fig. \ref{Hopfion_Plot_Triangular}). When we test it on configurations generated from two hopfions, it outputs $0$ with a good accuracy. If we test the neural network on mirror  reflected images of configurations obtained from one or two hopfions, it outputs a real number very close to the parity of the number of hopfions. Thus, the neural network tells us that on a stacked triangular lattice, hopfion is a topological defect characterized by $Z_2$ invariant, equal to the parity of the number of hopfions. Equivalently, we can conclude that on a non-bipartite lattice, the invariant  $\mathrm{sign} \: \mathrm{Pf}( M_{ij}n_{ij} )$ (discussed in the beginning of the Sec. \ref{Sec:CubicLattice}) is a physical topological invariant.



\section{Discussion}
\label{Sec:Discussion}

In this work, we have demonstrated that neural network can be used to distinguish topological phases of lattice dimer model. Using it, we verified that topological defects in the dimer model on a cubic lattice can be characterized by integer Hopf invariant, whereas on a stacked triangular lattice, the same defects are characterized by $Z_2$ invariant. 
In addition, in the case of the cubic lattice, we have explicitly checked 
 that 
topological defects are characterized by Hopf invariant
both in the case of open and periodic boundary conditions (strictly speaking, in the latter case, we limited our study to the subsector with trivial winding number). In fact, the neural network gave us a hint to the whole idea, that Hopf number can be defined on a lattice dimer model, which 
we verified analytically afterwards (see \cite{MyArticle}).  Thus, our paper can be viewed as the first work, where neural networks were successfully used to identify new topological phases. This is in contrast to the previous works, where machine learning algorithms were only able to identify previously known topological phases. 

We remark, that we found a qualitative difference between the optimal neural networks used in the cases of cubic and stacked triangular lattices. More specifically, we found that, in the case of cubic lattice, the neural network with just one hidden layer gives reasonably good predictions, whose accuracy increases with increasing number of hidden units. 
In contrast, in the case of stacked triangular lattice, one hidden layer is insufficient to train the neural network: the minimal required number of layers is three, 
 and the accuracy improves if we take larger number of hidden layers.
We suggest, that this difference may be related to complexity of the function, which the neural network approximates. Indeed, 
 on a cubic lattice, configurations are characterized by Hopf number, 
which can be expressed as a quadratic function of the effective magnetic field, or equivalently, dimer occupation number. However, 
on a stacked triangular lattice, the physical topological invariant is $\mathrm{sign} \mathrm{Pf}( M_{ij} n_{ij})$, which 
 is a high power function of the dimer occupation number. Thus, probing 3D lattice dimer model with neural networks leads us to conjecture, that the optimal number of hidden layers in a neural network is related to complexity of the function, which the neural network approximates. More generally, we believe that, in the future, it might be of interest to apply machine learning algorithms to simple physical systems in order to better understand the properties of machine learning algorithms themselves. 



We emphasize that, from our perspective, the main role of machine learning in physics is to provide insights about physical systems rather than rigorous results. In fact, like many other numerical methods, our approach has limitations due to fixed lattice size, finite number of samples in the datasets etc. For example, if one tries to draw conclusions based only on the neural networks, they may face such questions as e.g.: can it be that hopfions belong to a separate topological class on a $4\times 4 \times 4$ lattice, but to the same topological class on a larger lattice? Or, can it be that a hopfion does not belong to a separate topological class from a trivial configuration, but lies in a different part of the same topological class? The answers to these questions have to be found by using other techniques, than machine learning. Indeed, we claim that a hopfion is topologically distinct from a trivial dimer configuration (in both cases of cubic and stacked triangular lattices), because they have different values of a topological invariant $\mathrm{sign} \mathrm{Pf}( M_{ij} n_{ij})$. Furthermore, in Ref. \cite{MyArticle}, we claim that, in the case of cubic lattice, hopfions are characterized by an integer topological invariant based on its analytical derivation. On the other hand, our new result that on a stacked triangular lattice, $\mathrm{sign} \mathrm{Pf}( M_{ij} n_{ij})$ is a physical topological invariant, is suggestive - it has to be checked by other means. 

We mention that topological defects in lattice dimer model are interesting from physical point of view. In fact, 
lattice dimer models are known to be dual to various spin systems, many of which indeed have been realized experimentally. For example dimer model on a 3D diamond lattice 
(which, similarly to cubic, is also bipartite) is dual to spin ice on a pyrochlore lattice. The latter has been widely studied in the context of frustrated magnetism 
(see \cite{doi:10.1146/annurev-conmatphys-020911-125058, RevModPhys.82.53, 0034-4885-77-5-056501} for review),
and have been proven to exist in various materials, such as e.g.  $\mathrm{Dy_2 Ti_2 O_7}$ and $\mathrm{Ho_2 Ti_2 O_7}$. We believe that in these systems, it would be interesting to explore the effects resulting from the presence of distinct topological classes and consequently non-ergodicity.


We also suggest that hopfions can be experimentally realized in
 'artificial spin ice', where lattice dimer model is 
 simulated by nanomagnets (see e.g. \cite{RevModPhys.85.1473}). The simplest scenario of two-dimensional artificial spin ice has been extensively studied, and it 
has been found to share unique properties of lattice dimer model, such as magnetic monopoles and even Coulomb phase \cite{Perrin2016, Lao2018}. 
However, in the last years, there have been ongoing efforts to realize three-dimensional artificial spin ice \cite{FernGUndez-Pacheco2017, doi:10.1063/1.4861118, Keller2018}. Since, the existence of hopfions requires only two stacked 2D layers (see Fig. \ref{JustHopfion}), we expect that it should be possible to create hopfions in bilayer artificial spin ice, once it will become possible to realize dynamics through local flips only and to suppress other processes, e.g. monopole creation, longer loop flips etc. This will open wide opportunities both in the context of physics and quantum computing, because hopfions were predicted to host unique properties, such as e.g. non-Abelian anyons \cite{PhysRevLett.51.2250, PhysRevB.84.245119, PhysRevB.84.184501}.




\begin{acknowledgments}   
We would like to thank  Roger Melko 
for proposing this problem and having multiple discussions about it. We would also like to thank L. Sierens, B. Kulchitski , S. Wetzel, J. Rau, C. Nisoli, R. Moessner, L. Balents, L. Wang, A. Smith
for the discussions about this problem. 
Calculations were performed using the supercomputing facilities of Sharcnet.  
Financial support was provided by NSERC of Canada. 

\end{acknowledgments} 

\appendix
\section{Methods}
\label{Sec:Methods}

As we mentioned in the Sec. \ref{Sec:CubicLattice} we generate our training and testing datasets by applying local flips to configurations with fixed number of hopfions, shown on the Figs.  \ref{Hopfion_0}, \ref{Hopfion_1} and Figs. \ref{Hopfion_0}, \ref{Hopfion_1}, \ref{Hopfion_2} respectively. 
 We apply a thousand of random local flips between each pair of recorded configurations to make samples in the dataset more diverse. Samples with negative Hopf numbers are obtained by my mirror reflecting (over $100$ plane in the case of cubic lattice, and $001$ plane in the case of stacked triangular lattice) the samples with the positive Hopf numbers.

After generating the datasets, we train the neural networks. In the case of cubic lattice, we consider neural networks with one hidden and one outputs layers and vary the number of units in the hidden layer. In the case of triangular lattice, we vary the number of layers, but fix the number of units in each of them to be 128. In both cases, we use \textit{relu} as activation function in the hidden layers, and do not use any activation in the outputs layers. We perform our computations using Keras-2.1.5 library, 
 use SGD optimizer and minimize mean square error between the outputs and the Hopf numbers 0 and 1. This choice is natural assuming that the topological number can take any integer value. We perform a few hundreds of epochs and ensure that the loss decreases during training. 

Finally we test the neural network on $\sim 10^4$ configurations with each value of Hopf number. We compute the output of a fixed neural network for each configuration, and then compute its average and standard deviation. We present these results on the graphs \ref{HopfionPlots}, \ref{Hopfion_Plot_Periodic}, \ref{Hopfion_Plot_Triangular}.

As we mention in the main text, we use neural networks with one hidden layer in the case of the cubic lattice, and many hidden layers in the case of stacked triangular lattice. It is interesting to note, that in the case of cubic lattice, the neural network can distinguish Hopf numbers not used in training (i.e. "2", "-1", "-2") only if it has just one hidden layer. If we increase the number of layers, and test the neural network on configurations with Hopf number $2$, its output becomes closer to $1$, but if we test it on configurations with Hopf numbers $-1$ or $-2$, its output becomes close to $0$. We attribute this to overfitting, which naturally occurs if the number of training parameters is too large. We hypothesize that since Hopf number is a quadratic function of effective magnetic field, which is in turn proportional to a filling number, two layers are optimal for a neural network to distinguish it. In addition, we note that the accuracy of the neural networks decreases if we replace its activation functions with e.g. $tanh$ or $sigmoid$. We think that it happens, because the magnitudes of  such activation functions are constrained below one, and thus it is 'harder' to create a neural network configuration, which outputs 
a number close to an integer with a magnitude larger than one.

\section{A brief construction of the Hopf number}
\label{Sec:HopfNumber}

In this section we briefly describe analytical derivation of the Hopf invariant in 3D lattice dimer model. As we mentioned previously, Hopf invariant can be defined on a bipartite lattice, i.e. whose sites can be labeled by $\sigma = \pm 1$ in such a way that any bond connects sites with opposite values of $\sigma$. A particular example of a bipartite lattice is cubic lattice, whereas an example of a non-bipartite lattice is stacked triangular lattice.  
The main idea is that on a bipartite lattice, dimer filling can be described in terms of effective magnetic field
\begin{eqnarray}
B_{i} (\vec{r}) = \sigma (n_{r , r+e_i} - w_{r , r+e_i} ).
\nonumber
\end{eqnarray}
 Here $w_{r , r+e_i}$ is a fixed number characterizing each bond of the lattice. Conventionally, $w_{r , r+e_i}$ is chosen to be equal to inverse coordination number of the lattice, but in Ref. \cite{MyArticle}, we develope a more general approach, which makes it possible to account for finite size of the lattice.  As one can see, $B_{i} (\vec{r})$ characterizes filling of each bond connecting the sites at positions $\vec{r}$ and $\vec{r} + \vec{e_i}$.  It is called an effective magnetic field, because it satisfies the constraint of zero divergence:
\begin{eqnarray}
&&B_x(x, y, z) - B_x(x-1, y, z)
\nonumber\\
&&+ B_y(x, y, z) - B_y(x, y-1, z)
\nonumber\\
&&+ B_z(x, y, z) - B_z(x, y, z-1)
=0,
\label{DivB}
\end{eqnarray}
which in turn follows from the constraint of exactly one dimer attached to each site. 

The condition of zero divergence (\ref{DivB}) makes it possible to introduce effective vector potential - a vector defined at each plaquette, such that its discrete rotor gives the effective magnetic field
\begin{eqnarray}
&& B_x (x,y,z) = A_z (x,y,z) - A_z(x,y-1,z)
\nonumber\\
&& \qquad \qquad \qquad - A_y(x,y,z) + A_y (x,y,z-1),
\nonumber\\
&& \quad  B_{y,z} \:\mbox{  are defined through cyclic permutations}.
\nonumber
\end{eqnarray}

Once we defined the effective magnetic field and vector potential, we can write Hopf number as a sum of their products
\begin{eqnarray}
\chi &=& \sum\limits_{x,y,z} 
\frac{A_x (x,y,z) }{8} 
\label{DiscreteHopfNumber}
\\
& \times &
\left( 
B_x(x,y,z) + B_x (x,y+1,z) 
\right. \nonumber\\ && \left.
+ B_x (x,y,z+1) + B_x (x,y+1,z+1)
\right. \nonumber\\ && \left.
+ B_x(x-1,y,z) + B_x (x-1,y+1,z) 
\right. \nonumber\\ && \left.
+ B_x (x-1,y,z+1) + B_x (x-1,y+1,z+1)
\right) 
\nonumber\\
&+&( \mbox{cyclic permutations}).
\nonumber
\end{eqnarray}
Here we took vector potentials at each plaquette and multiplied it by average magentic field along the bonds adjacent and perpendicular to the plaquette. 

Through explicit calculations, (see Ref. \cite{MyArticle}) one may check that the Hopf number (\ref{DiscreteHopfNumber}) is
\begin{itemize} \itemsep2pt
\item gauge-invariant
\item invariant under local dimer flips
\item is equal to $0$, $1$, $2$ for the dimer configurations shown on the Figs. \ref{Hopfion_0}, \ref{Hopfion_1}, \ref{Hopfion_2} correspondingly and changes sign through mirror reflection. 
\end{itemize}

Thus, the result about about existence of Hopf number in 3D lattice dimer model, which was first obtained by using neural networks, can be verified analytically.


\bibliographystyle{apsrev4-1}
\bibliography{HopfLinksBib}

\begin{thebibliography}{39}%
\makeatletter
\providecommand \@ifxundefined [1]{%
 \@ifx{#1\undefined}
}%
\providecommand \@ifnum [1]{%
 \ifnum #1\expandafter \@firstoftwo
 \else \expandafter \@secondoftwo
 \fi
}%
\providecommand \@ifx [1]{%
 \ifx #1\expandafter \@firstoftwo
 \else \expandafter \@secondoftwo
 \fi
}%
\providecommand \natexlab [1]{#1}%
\providecommand \enquote  [1]{``#1''}%
\providecommand \bibnamefont  [1]{#1}%
\providecommand \bibfnamefont [1]{#1}%
\providecommand \citenamefont [1]{#1}%
\providecommand \href@noop [0]{\@secondoftwo}%
\providecommand \href [0]{\begingroup \@sanitize@url \@href}%
\providecommand \@href[1]{\@@startlink{#1}\@@href}%
\providecommand \@@href[1]{\endgroup#1\@@endlink}%
\providecommand \@sanitize@url [0]{\catcode `\\12\catcode `\$12\catcode
  `\&12\catcode `\#12\catcode `\^12\catcode `\_12\catcode `\%12\relax}%
\providecommand \@@startlink[1]{}%
\providecommand \@@endlink[0]{}%
\providecommand \url  [0]{\begingroup\@sanitize@url \@url }%
\providecommand \@url [1]{\endgroup\@href {#1}{\urlprefix }}%
\providecommand \urlprefix  [0]{URL }%
\providecommand \Eprint [0]{\href }%
\providecommand \doibase [0]{http://dx.doi.org/}%
\providecommand \selectlanguage [0]{\@gobble}%
\providecommand \bibinfo  [0]{\@secondoftwo}%
\providecommand \bibfield  [0]{\@secondoftwo}%
\providecommand \translation [1]{[#1]}%
\providecommand \BibitemOpen [0]{}%
\providecommand \bibitemStop [0]{}%
\providecommand \bibitemNoStop [0]{.\EOS\space}%
\providecommand \EOS [0]{\spacefactor3000\relax}%
\providecommand \BibitemShut  [1]{\csname bibitem#1\endcsname}%
\let\auto@bib@innerbib\@empty
\bibitem [{\citenamefont {James}\ \emph {et~al.}(2014)\citenamefont {James},
  \citenamefont {Witten}, \citenamefont {Hastie},\ and\ \citenamefont
  {Tibshirani}}]{james2014introduction}%
  \BibitemOpen
  \bibfield  {author} {\bibinfo {author} {\bibfnamefont {G.}~\bibnamefont
  {James}}, \bibinfo {author} {\bibfnamefont {D.}~\bibnamefont {Witten}},
  \bibinfo {author} {\bibfnamefont {T.}~\bibnamefont {Hastie}}, \ and\ \bibinfo
  {author} {\bibfnamefont {R.}~\bibnamefont {Tibshirani}},\ }\href
  {https://books.google.com/books?id=at1bmAEACAAJ} {\emph {\bibinfo {title} {An
  Introduction to Statistical Learning: with Applications in R}}},\ Springer
  Texts in Statistics\ (\bibinfo  {publisher} {Springer New York},\ \bibinfo
  {year} {2014})\BibitemShut {NoStop}%
\bibitem [{\citenamefont {Goodfellow}\ \emph {et~al.}(2016)\citenamefont
  {Goodfellow}, \citenamefont {Bengio},\ and\ \citenamefont
  {Courville}}]{Goodfellow-et-al-2016}%
  \BibitemOpen
  \bibfield  {author} {\bibinfo {author} {\bibfnamefont {I.}~\bibnamefont
  {Goodfellow}}, \bibinfo {author} {\bibfnamefont {Y.}~\bibnamefont {Bengio}},
  \ and\ \bibinfo {author} {\bibfnamefont {A.}~\bibnamefont {Courville}},\
  }\href@noop {} {\emph {\bibinfo {title} {Deep Learning}}}\ (\bibinfo
  {publisher} {MIT Press},\ \bibinfo {year} {2016})\ \bibinfo {note}
  {\url{http://www.deeplearningbook.org}}\BibitemShut {NoStop}%
\bibitem [{\citenamefont {Carrasquilla}\ and\ \citenamefont
  {Melko}(2017)}]{Carrasquilla2017}%
  \BibitemOpen
  \bibfield  {author} {\bibinfo {author} {\bibfnamefont {J.}~\bibnamefont
  {Carrasquilla}}\ and\ \bibinfo {author} {\bibfnamefont {R.~G.}\ \bibnamefont
  {Melko}},\ }\href {http://dx.doi.org/10.1038/nphys4035} {\bibfield  {journal}
  {\bibinfo  {journal} {Nature Physics}\ }\textbf {\bibinfo {volume} {13}},\
  \bibinfo {pages} {431 EP } (\bibinfo {year} {2017})}\BibitemShut {NoStop}%
\bibitem [{\citenamefont {Ch'ng}\ \emph {et~al.}(2017)\citenamefont {Ch'ng},
  \citenamefont {Carrasquilla}, \citenamefont {Melko},\ and\ \citenamefont
  {Khatami}}]{PhysRevX.7.031038}%
  \BibitemOpen
  \bibfield  {author} {\bibinfo {author} {\bibfnamefont {K.}~\bibnamefont
  {Ch'ng}}, \bibinfo {author} {\bibfnamefont {J.}~\bibnamefont {Carrasquilla}},
  \bibinfo {author} {\bibfnamefont {R.~G.}\ \bibnamefont {Melko}}, \ and\
  \bibinfo {author} {\bibfnamefont {E.}~\bibnamefont {Khatami}},\ }\href
  {\doibase 10.1103/PhysRevX.7.031038} {\bibfield  {journal} {\bibinfo
  {journal} {Phys. Rev. X}\ }\textbf {\bibinfo {volume} {7}},\ \bibinfo {pages}
  {031038} (\bibinfo {year} {2017})}\BibitemShut {NoStop}%
\bibitem [{\citenamefont {Zhang}\ \emph {et~al.}(2018)\citenamefont {Zhang},
  \citenamefont {Shen},\ and\ \citenamefont {Zhai}}]{PhysRevLett.120.066401}%
  \BibitemOpen
  \bibfield  {author} {\bibinfo {author} {\bibfnamefont {P.}~\bibnamefont
  {Zhang}}, \bibinfo {author} {\bibfnamefont {H.}~\bibnamefont {Shen}}, \ and\
  \bibinfo {author} {\bibfnamefont {H.}~\bibnamefont {Zhai}},\ }\href {\doibase
  10.1103/PhysRevLett.120.066401} {\bibfield  {journal} {\bibinfo  {journal}
  {Phys. Rev. Lett.}\ }\textbf {\bibinfo {volume} {120}},\ \bibinfo {pages}
  {066401} (\bibinfo {year} {2018})}\BibitemShut {NoStop}%
\bibitem [{\citenamefont {Zhang}\ \emph {et~al.}(2017)\citenamefont {Zhang},
  \citenamefont {Melko},\ and\ \citenamefont {Kim}}]{zhang2017machine}%
  \BibitemOpen
  \bibfield  {author} {\bibinfo {author} {\bibfnamefont {Y.}~\bibnamefont
  {Zhang}}, \bibinfo {author} {\bibfnamefont {R.~G.}\ \bibnamefont {Melko}}, \
  and\ \bibinfo {author} {\bibfnamefont {E.-A.}\ \bibnamefont {Kim}},\
  }\href@noop {} {\bibfield  {journal} {\bibinfo  {journal} {Physical Review
  B}\ }\textbf {\bibinfo {volume} {96}},\ \bibinfo {pages} {245119} (\bibinfo
  {year} {2017})}\BibitemShut {NoStop}%
\bibitem [{\citenamefont {Beach}\ \emph {et~al.}(2018)\citenamefont {Beach},
  \citenamefont {Golubeva},\ and\ \citenamefont {Melko}}]{beach2018machine}%
  \BibitemOpen
  \bibfield  {author} {\bibinfo {author} {\bibfnamefont {M.~J.}\ \bibnamefont
  {Beach}}, \bibinfo {author} {\bibfnamefont {A.}~\bibnamefont {Golubeva}}, \
  and\ \bibinfo {author} {\bibfnamefont {R.~G.}\ \bibnamefont {Melko}},\
  }\href@noop {} {\bibfield  {journal} {\bibinfo  {journal} {Physical Review
  B}\ }\textbf {\bibinfo {volume} {97}},\ \bibinfo {pages} {045207} (\bibinfo
  {year} {2018})}\BibitemShut {NoStop}%
\bibitem [{\citenamefont {Venderley}\ \emph {et~al.}(2018)\citenamefont
  {Venderley}, \citenamefont {Khemani},\ and\ \citenamefont
  {Kim}}]{PhysRevLett.120.257204}%
  \BibitemOpen
  \bibfield  {author} {\bibinfo {author} {\bibfnamefont {J.}~\bibnamefont
  {Venderley}}, \bibinfo {author} {\bibfnamefont {V.}~\bibnamefont {Khemani}},
  \ and\ \bibinfo {author} {\bibfnamefont {E.-A.}\ \bibnamefont {Kim}},\ }\href
  {\doibase 10.1103/PhysRevLett.120.257204} {\bibfield  {journal} {\bibinfo
  {journal} {Phys. Rev. Lett.}\ }\textbf {\bibinfo {volume} {120}},\ \bibinfo
  {pages} {257204} (\bibinfo {year} {2018})}\BibitemShut {NoStop}%
\bibitem [{\citenamefont {Deng}\ \emph {et~al.}(2017)\citenamefont {Deng},
  \citenamefont {Li},\ and\ \citenamefont {Das~Sarma}}]{PhysRevB.96.195145}%
  \BibitemOpen
  \bibfield  {author} {\bibinfo {author} {\bibfnamefont {D.-L.}\ \bibnamefont
  {Deng}}, \bibinfo {author} {\bibfnamefont {X.}~\bibnamefont {Li}}, \ and\
  \bibinfo {author} {\bibfnamefont {S.}~\bibnamefont {Das~Sarma}},\ }\href
  {\doibase 10.1103/PhysRevB.96.195145} {\bibfield  {journal} {\bibinfo
  {journal} {Phys. Rev. B}\ }\textbf {\bibinfo {volume} {96}},\ \bibinfo
  {pages} {195145} (\bibinfo {year} {2017})}\BibitemShut {NoStop}%
\bibitem [{\citenamefont {Torlai}\ \emph {et~al.}(2018)\citenamefont {Torlai},
  \citenamefont {Mazzola}, \citenamefont {Carrasquilla}, \citenamefont
  {Troyer}, \citenamefont {Melko},\ and\ \citenamefont {Carleo}}]{Torlai2018}%
  \BibitemOpen
  \bibfield  {author} {\bibinfo {author} {\bibfnamefont {G.}~\bibnamefont
  {Torlai}}, \bibinfo {author} {\bibfnamefont {G.}~\bibnamefont {Mazzola}},
  \bibinfo {author} {\bibfnamefont {J.}~\bibnamefont {Carrasquilla}}, \bibinfo
  {author} {\bibfnamefont {M.}~\bibnamefont {Troyer}}, \bibinfo {author}
  {\bibfnamefont {R.}~\bibnamefont {Melko}}, \ and\ \bibinfo {author}
  {\bibfnamefont {G.}~\bibnamefont {Carleo}},\ }\href {\doibase
  10.1038/s41567-018-0048-5} {\bibfield  {journal} {\bibinfo  {journal} {Nature
  Physics}\ }\textbf {\bibinfo {volume} {14}},\ \bibinfo {pages} {447}
  (\bibinfo {year} {2018})}\BibitemShut {NoStop}%
\bibitem [{\citenamefont {Kasteleyn}(1963)}]{doi:10.1063/1.1703953}%
  \BibitemOpen
  \bibfield  {author} {\bibinfo {author} {\bibfnamefont {P.~W.}\ \bibnamefont
  {Kasteleyn}},\ }\href {\doibase 10.1063/1.1703953} {\bibfield  {journal}
  {\bibinfo  {journal} {Journal of Mathematical Physics}\ }\textbf {\bibinfo
  {volume} {4}},\ \bibinfo {pages} {287} (\bibinfo {year} {1963})},\ \Eprint
  {http://arxiv.org/abs/https://doi.org/10.1063/1.1703953}
  {https://doi.org/10.1063/1.1703953} \BibitemShut {NoStop}%
\bibitem [{\citenamefont {Fisher}(1966)}]{doi:10.1063/1.1704825}%
  \BibitemOpen
  \bibfield  {author} {\bibinfo {author} {\bibfnamefont {M.~E.}\ \bibnamefont
  {Fisher}},\ }\href {\doibase 10.1063/1.1704825} {\bibfield  {journal}
  {\bibinfo  {journal} {Journal of Mathematical Physics}\ }\textbf {\bibinfo
  {volume} {7}},\ \bibinfo {pages} {1776} (\bibinfo {year} {1966})},\ \Eprint
  {http://arxiv.org/abs/https://doi.org/10.1063/1.1704825}
  {https://doi.org/10.1063/1.1704825} \BibitemShut {NoStop}%
\bibitem [{\citenamefont {ANDERSON}(1987)}]{ANDERSON1196}%
  \BibitemOpen
  \bibfield  {author} {\bibinfo {author} {\bibfnamefont {P.~W.}\ \bibnamefont
  {ANDERSON}},\ }\href {\doibase 10.1126/science.235.4793.1196} {\bibfield
  {journal} {\bibinfo  {journal} {Science}\ }\textbf {\bibinfo {volume}
  {235}},\ \bibinfo {pages} {1196} (\bibinfo {year} {1987})}\BibitemShut
  {NoStop}%
\bibitem [{\citenamefont {Rokhsar}\ and\ \citenamefont
  {Kivelson}(1988)}]{PhysRevLett.61.2376}%
  \BibitemOpen
  \bibfield  {author} {\bibinfo {author} {\bibfnamefont {D.~S.}\ \bibnamefont
  {Rokhsar}}\ and\ \bibinfo {author} {\bibfnamefont {S.~A.}\ \bibnamefont
  {Kivelson}},\ }\href {\doibase 10.1103/PhysRevLett.61.2376} {\bibfield
  {journal} {\bibinfo  {journal} {Phys. Rev. Lett.}\ }\textbf {\bibinfo
  {volume} {61}},\ \bibinfo {pages} {2376} (\bibinfo {year}
  {1988})}\BibitemShut {NoStop}%
\bibitem [{\citenamefont {Sikora}\ \emph {et~al.}(2011)\citenamefont {Sikora},
  \citenamefont {Shannon}, \citenamefont {Pollmann}, \citenamefont {Penc},\
  and\ \citenamefont {Fulde}}]{PhysRevB.84.115129}%
  \BibitemOpen
  \bibfield  {author} {\bibinfo {author} {\bibfnamefont {O.}~\bibnamefont
  {Sikora}}, \bibinfo {author} {\bibfnamefont {N.}~\bibnamefont {Shannon}},
  \bibinfo {author} {\bibfnamefont {F.}~\bibnamefont {Pollmann}}, \bibinfo
  {author} {\bibfnamefont {K.}~\bibnamefont {Penc}}, \ and\ \bibinfo {author}
  {\bibfnamefont {P.}~\bibnamefont {Fulde}},\ }\href {\doibase
  10.1103/PhysRevB.84.115129} {\bibfield  {journal} {\bibinfo  {journal} {Phys.
  Rev. B}\ }\textbf {\bibinfo {volume} {84}},\ \bibinfo {pages} {115129}
  (\bibinfo {year} {2011})}\BibitemShut {NoStop}%
\bibitem [{\citenamefont {Hermele}\ \emph {et~al.}(2004)\citenamefont
  {Hermele}, \citenamefont {Fisher},\ and\ \citenamefont
  {Balents}}]{PhysRevB.69.064404}%
  \BibitemOpen
  \bibfield  {author} {\bibinfo {author} {\bibfnamefont {M.}~\bibnamefont
  {Hermele}}, \bibinfo {author} {\bibfnamefont {M.~P.~A.}\ \bibnamefont
  {Fisher}}, \ and\ \bibinfo {author} {\bibfnamefont {L.}~\bibnamefont
  {Balents}},\ }\href {\doibase 10.1103/PhysRevB.69.064404} {\bibfield
  {journal} {\bibinfo  {journal} {Phys. Rev. B}\ }\textbf {\bibinfo {volume}
  {69}},\ \bibinfo {pages} {064404} (\bibinfo {year} {2004})}\BibitemShut
  {NoStop}%
\bibitem [{\citenamefont {Castelnovo}\ \emph {et~al.}(2012)\citenamefont
  {Castelnovo}, \citenamefont {Moessner},\ and\ \citenamefont
  {Sondhi}}]{doi:10.1146/annurev-conmatphys-020911-125058}%
  \BibitemOpen
  \bibfield  {author} {\bibinfo {author} {\bibfnamefont {C.}~\bibnamefont
  {Castelnovo}}, \bibinfo {author} {\bibfnamefont {R.}~\bibnamefont
  {Moessner}}, \ and\ \bibinfo {author} {\bibfnamefont {S.}~\bibnamefont
  {Sondhi}},\ }\href {\doibase 10.1146/annurev-conmatphys-020911-125058}
  {\bibfield  {journal} {\bibinfo  {journal} {Annual Review of Condensed Matter
  Physics}\ }\textbf {\bibinfo {volume} {3}},\ \bibinfo {pages} {35} (\bibinfo
  {year} {2012})},\ \Eprint
  {http://arxiv.org/abs/https://doi.org/10.1146/annurev-conmatphys-020911-125058}
  {https://doi.org/10.1146/annurev-conmatphys-020911-125058} \BibitemShut
  {NoStop}%
\bibitem [{\citenamefont {Gardner}\ \emph {et~al.}(2010)\citenamefont
  {Gardner}, \citenamefont {Gingras},\ and\ \citenamefont
  {Greedan}}]{RevModPhys.82.53}%
  \BibitemOpen
  \bibfield  {author} {\bibinfo {author} {\bibfnamefont {J.~S.}\ \bibnamefont
  {Gardner}}, \bibinfo {author} {\bibfnamefont {M.~J.~P.}\ \bibnamefont
  {Gingras}}, \ and\ \bibinfo {author} {\bibfnamefont {J.~E.}\ \bibnamefont
  {Greedan}},\ }\href {\doibase 10.1103/RevModPhys.82.53} {\bibfield  {journal}
  {\bibinfo  {journal} {Rev. Mod. Phys.}\ }\textbf {\bibinfo {volume} {82}},\
  \bibinfo {pages} {53} (\bibinfo {year} {2010})}\BibitemShut {NoStop}%
\bibitem [{\citenamefont {Nisoli}\ \emph {et~al.}(2013)\citenamefont {Nisoli},
  \citenamefont {Moessner},\ and\ \citenamefont
  {Schiffer}}]{RevModPhys.85.1473}%
  \BibitemOpen
  \bibfield  {author} {\bibinfo {author} {\bibfnamefont {C.}~\bibnamefont
  {Nisoli}}, \bibinfo {author} {\bibfnamefont {R.}~\bibnamefont {Moessner}}, \
  and\ \bibinfo {author} {\bibfnamefont {P.}~\bibnamefont {Schiffer}},\ }\href
  {\doibase 10.1103/RevModPhys.85.1473} {\bibfield  {journal} {\bibinfo
  {journal} {Rev. Mod. Phys.}\ }\textbf {\bibinfo {volume} {85}},\ \bibinfo
  {pages} {1473} (\bibinfo {year} {2013})}\BibitemShut {NoStop}%
\bibitem [{\citenamefont {Perrin}\ \emph {et~al.}(2016)\citenamefont {Perrin},
  \citenamefont {Canals},\ and\ \citenamefont {Rougemaille}}]{Perrin2016}%
  \BibitemOpen
  \bibfield  {author} {\bibinfo {author} {\bibfnamefont {Y.}~\bibnamefont
  {Perrin}}, \bibinfo {author} {\bibfnamefont {B.}~\bibnamefont {Canals}}, \
  and\ \bibinfo {author} {\bibfnamefont {N.}~\bibnamefont {Rougemaille}},\
  }\href {http://dx.doi.org/10.1038/nature20155} {\bibfield  {journal}
  {\bibinfo  {journal} {Nature}\ }\textbf {\bibinfo {volume} {540}},\ \bibinfo
  {pages} {410 EP } (\bibinfo {year} {2016})}\BibitemShut {NoStop}%
\bibitem [{\citenamefont {Lao}\ \emph {et~al.}(2018)\citenamefont {Lao},
  \citenamefont {Caravelli}, \citenamefont {Sheikh}, \citenamefont {Sklenar},
  \citenamefont {Gardeazabal}, \citenamefont {Watts}, \citenamefont {Albrecht},
  \citenamefont {Scholl}, \citenamefont {Dahmen}, \citenamefont {Nisoli},\ and\
  \citenamefont {Schiffer}}]{Lao2018}%
  \BibitemOpen
  \bibfield  {author} {\bibinfo {author} {\bibfnamefont {Y.}~\bibnamefont
  {Lao}}, \bibinfo {author} {\bibfnamefont {F.}~\bibnamefont {Caravelli}},
  \bibinfo {author} {\bibfnamefont {M.}~\bibnamefont {Sheikh}}, \bibinfo
  {author} {\bibfnamefont {J.}~\bibnamefont {Sklenar}}, \bibinfo {author}
  {\bibfnamefont {D.}~\bibnamefont {Gardeazabal}}, \bibinfo {author}
  {\bibfnamefont {J.~D.}\ \bibnamefont {Watts}}, \bibinfo {author}
  {\bibfnamefont {A.~M.}\ \bibnamefont {Albrecht}}, \bibinfo {author}
  {\bibfnamefont {A.}~\bibnamefont {Scholl}}, \bibinfo {author} {\bibfnamefont
  {K.}~\bibnamefont {Dahmen}}, \bibinfo {author} {\bibfnamefont
  {C.}~\bibnamefont {Nisoli}}, \ and\ \bibinfo {author} {\bibfnamefont
  {P.}~\bibnamefont {Schiffer}},\ }\href {\doibase 10.1038/s41567-018-0077-0}
  {\bibfield  {journal} {\bibinfo  {journal} {Nature Physics}\ }\textbf
  {\bibinfo {volume} {14}},\ \bibinfo {pages} {723} (\bibinfo {year}
  {2018})}\BibitemShut {NoStop}%
\bibitem [{\citenamefont {Chern}\ \emph {et~al.}(2014)\citenamefont {Chern},
  \citenamefont {Reichhardt},\ and\ \citenamefont
  {Nisoli}}]{doi:10.1063/1.4861118}%
  \BibitemOpen
  \bibfield  {author} {\bibinfo {author} {\bibfnamefont {G.-W.}\ \bibnamefont
  {Chern}}, \bibinfo {author} {\bibfnamefont {C.}~\bibnamefont {Reichhardt}}, \
  and\ \bibinfo {author} {\bibfnamefont {C.}~\bibnamefont {Nisoli}},\ }\href
  {\doibase 10.1063/1.4861118} {\bibfield  {journal} {\bibinfo  {journal}
  {Applied Physics Letters}\ }\textbf {\bibinfo {volume} {104}},\ \bibinfo
  {pages} {013101} (\bibinfo {year} {2014})},\ \Eprint
  {http://arxiv.org/abs/https://doi.org/10.1063/1.4861118}
  {https://doi.org/10.1063/1.4861118} \BibitemShut {NoStop}%
\bibitem [{\citenamefont {Keller}\ \emph {et~al.}(2018)\citenamefont {Keller},
  \citenamefont {Al~Mamoori}, \citenamefont {Pieper}, \citenamefont {Gspan},
  \citenamefont {Stockem}, \citenamefont {Schroder}, \citenamefont {Barth},
  \citenamefont {Winkler}, \citenamefont {Plank}, \citenamefont {Pohlit},
  \citenamefont {MGjller},\ and\ \citenamefont {Huth}}]{Keller2018}%
  \BibitemOpen
  \bibfield  {author} {\bibinfo {author} {\bibfnamefont {L.}~\bibnamefont
  {Keller}}, \bibinfo {author} {\bibfnamefont {M.~K.~I.}\ \bibnamefont
  {Al~Mamoori}}, \bibinfo {author} {\bibfnamefont {J.}~\bibnamefont {Pieper}},
  \bibinfo {author} {\bibfnamefont {C.}~\bibnamefont {Gspan}}, \bibinfo
  {author} {\bibfnamefont {I.}~\bibnamefont {Stockem}}, \bibinfo {author}
  {\bibfnamefont {C.}~\bibnamefont {Schroder}}, \bibinfo {author}
  {\bibfnamefont {S.}~\bibnamefont {Barth}}, \bibinfo {author} {\bibfnamefont
  {R.}~\bibnamefont {Winkler}}, \bibinfo {author} {\bibfnamefont
  {H.}~\bibnamefont {Plank}}, \bibinfo {author} {\bibfnamefont
  {M.}~\bibnamefont {Pohlit}}, \bibinfo {author} {\bibfnamefont
  {J.}~\bibnamefont {MGjller}}, \ and\ \bibinfo {author} {\bibfnamefont
  {M.}~\bibnamefont {Huth}},\ }\href {\doibase 10.1038/s41598-018-24431-x}
  {\bibfield  {journal} {\bibinfo  {journal} {Scientific Reports}\ }\textbf
  {\bibinfo {volume} {8}},\ \bibinfo {pages} {6160} (\bibinfo {year}
  {2018})}\BibitemShut {NoStop}%
\bibitem [{\citenamefont {Moessner}\ and\ \citenamefont
  {Sondhi}(2003)}]{PhysRevB.68.184512}%
  \BibitemOpen
  \bibfield  {author} {\bibinfo {author} {\bibfnamefont {R.}~\bibnamefont
  {Moessner}}\ and\ \bibinfo {author} {\bibfnamefont {S.~L.}\ \bibnamefont
  {Sondhi}},\ }\href {\doibase 10.1103/PhysRevB.68.184512} {\bibfield
  {journal} {\bibinfo  {journal} {Phys. Rev. B}\ }\textbf {\bibinfo {volume}
  {68}},\ \bibinfo {pages} {184512} (\bibinfo {year} {2003})}\BibitemShut
  {NoStop}%
\bibitem [{\citenamefont {Huse}\ \emph {et~al.}(2003)\citenamefont {Huse},
  \citenamefont {Krauth}, \citenamefont {Moessner},\ and\ \citenamefont
  {Sondhi}}]{PhysRevLett.91.167004}%
  \BibitemOpen
  \bibfield  {author} {\bibinfo {author} {\bibfnamefont {D.~A.}\ \bibnamefont
  {Huse}}, \bibinfo {author} {\bibfnamefont {W.}~\bibnamefont {Krauth}},
  \bibinfo {author} {\bibfnamefont {R.}~\bibnamefont {Moessner}}, \ and\
  \bibinfo {author} {\bibfnamefont {S.~L.}\ \bibnamefont {Sondhi}},\ }\href
  {\doibase 10.1103/PhysRevLett.91.167004} {\bibfield  {journal} {\bibinfo
  {journal} {Phys. Rev. Lett.}\ }\textbf {\bibinfo {volume} {91}},\ \bibinfo
  {pages} {167004} (\bibinfo {year} {2003})}\BibitemShut {NoStop}%
\bibitem [{\citenamefont {Freedman}\ \emph {et~al.}(2011)\citenamefont
  {Freedman}, \citenamefont {Hastings}, \citenamefont {Nayak},\ and\
  \citenamefont {Qi}}]{PhysRevB.84.245119}%
  \BibitemOpen
  \bibfield  {author} {\bibinfo {author} {\bibfnamefont {M.}~\bibnamefont
  {Freedman}}, \bibinfo {author} {\bibfnamefont {M.~B.}\ \bibnamefont
  {Hastings}}, \bibinfo {author} {\bibfnamefont {C.}~\bibnamefont {Nayak}}, \
  and\ \bibinfo {author} {\bibfnamefont {X.-L.}\ \bibnamefont {Qi}},\ }\href
  {\doibase 10.1103/PhysRevB.84.245119} {\bibfield  {journal} {\bibinfo
  {journal} {Phys. Rev. B}\ }\textbf {\bibinfo {volume} {84}},\ \bibinfo
  {pages} {245119} (\bibinfo {year} {2011})}\BibitemShut {NoStop}%
\bibitem [{\citenamefont {E}(1996)}]{e1996force}%
  \BibitemOpen
  \bibfield  {author} {\bibinfo {author} {\bibfnamefont {M.}~\bibnamefont
  {E}},\ }\href {https://books.google.ca/books?id=8fHsCgAAQBAJ} {\emph
  {\bibinfo {title} {Force-free Magnetic Fields: Solutions, Topology And
  Applications}}}\ (\bibinfo  {publisher} {World Scientific Publishing
  Company},\ \bibinfo {year} {1996})\BibitemShut {NoStop}%
\bibitem [{\citenamefont {Arnold}\ and\ \citenamefont
  {Khesin}(2013)}]{arnold2013topological}%
  \BibitemOpen
  \bibfield  {author} {\bibinfo {author} {\bibfnamefont {V.}~\bibnamefont
  {Arnold}}\ and\ \bibinfo {author} {\bibfnamefont {B.}~\bibnamefont
  {Khesin}},\ }\href {https://books.google.ca/books?id=NsA0ngEACAAJ} {\emph
  {\bibinfo {title} {Topological Properties of Magnetic and Vorticity Fields.
  In: Topological Methods in Hydrodynamics}}},\ Applied Mathematical Sciences,
  vol 125.\ (\bibinfo  {publisher} {Springer New York},\ \bibinfo {year}
  {2013})\BibitemShut {NoStop}%
\bibitem [{\citenamefont {Ran}\ \emph {et~al.}(2011)\citenamefont {Ran},
  \citenamefont {Hosur},\ and\ \citenamefont
  {Vishwanath}}]{PhysRevB.84.184501}%
  \BibitemOpen
  \bibfield  {author} {\bibinfo {author} {\bibfnamefont {Y.}~\bibnamefont
  {Ran}}, \bibinfo {author} {\bibfnamefont {P.}~\bibnamefont {Hosur}}, \ and\
  \bibinfo {author} {\bibfnamefont {A.}~\bibnamefont {Vishwanath}},\ }\href
  {\doibase 10.1103/PhysRevB.84.184501} {\bibfield  {journal} {\bibinfo
  {journal} {Phys. Rev. B}\ }\textbf {\bibinfo {volume} {84}},\ \bibinfo
  {pages} {184501} (\bibinfo {year} {2011})}\BibitemShut {NoStop}%
\bibitem [{\citenamefont {Wilczek}\ and\ \citenamefont
  {Zee}(1983)}]{PhysRevLett.51.2250}%
  \BibitemOpen
  \bibfield  {author} {\bibinfo {author} {\bibfnamefont {F.}~\bibnamefont
  {Wilczek}}\ and\ \bibinfo {author} {\bibfnamefont {A.}~\bibnamefont {Zee}},\
  }\href {\doibase 10.1103/PhysRevLett.51.2250} {\bibfield  {journal} {\bibinfo
   {journal} {Phys. Rev. Lett.}\ }\textbf {\bibinfo {volume} {51}},\ \bibinfo
  {pages} {2250} (\bibinfo {year} {1983})}\BibitemShut {NoStop}%
\bibitem [{\citenamefont {Deng}\ \emph {et~al.}(2013)\citenamefont {Deng},
  \citenamefont {Wang}, \citenamefont {Shen},\ and\ \citenamefont
  {Duan}}]{PhysRevB.88.201105}%
  \BibitemOpen
  \bibfield  {author} {\bibinfo {author} {\bibfnamefont {D.-L.}\ \bibnamefont
  {Deng}}, \bibinfo {author} {\bibfnamefont {S.-T.}\ \bibnamefont {Wang}},
  \bibinfo {author} {\bibfnamefont {C.}~\bibnamefont {Shen}}, \ and\ \bibinfo
  {author} {\bibfnamefont {L.-M.}\ \bibnamefont {Duan}},\ }\href {\doibase
  10.1103/PhysRevB.88.201105} {\bibfield  {journal} {\bibinfo  {journal} {Phys.
  Rev. B}\ }\textbf {\bibinfo {volume} {88}},\ \bibinfo {pages} {201105(R)}
  (\bibinfo {year} {2013})}\BibitemShut {NoStop}%
\bibitem [{\citenamefont {Kennedy}(2016)}]{PhysRevB.94.035137}%
  \BibitemOpen
  \bibfield  {author} {\bibinfo {author} {\bibfnamefont {R.}~\bibnamefont
  {Kennedy}},\ }\href {\doibase 10.1103/PhysRevB.94.035137} {\bibfield
  {journal} {\bibinfo  {journal} {Phys. Rev. B}\ }\textbf {\bibinfo {volume}
  {94}},\ \bibinfo {pages} {035137} (\bibinfo {year} {2016})}\BibitemShut
  {NoStop}%
\bibitem [{\citenamefont {Deng}\ \emph {et~al.}(2018)\citenamefont {Deng},
  \citenamefont {Wang}, \citenamefont {Sun},\ and\ \citenamefont
  {Duan}}]{0256-307X-35-1-013701}%
  \BibitemOpen
  \bibfield  {author} {\bibinfo {author} {\bibfnamefont {D.-L.}\ \bibnamefont
  {Deng}}, \bibinfo {author} {\bibfnamefont {S.-T.}\ \bibnamefont {Wang}},
  \bibinfo {author} {\bibfnamefont {K.}~\bibnamefont {Sun}}, \ and\ \bibinfo
  {author} {\bibfnamefont {L.-M.}\ \bibnamefont {Duan}},\ }\href
  {http://stacks.iop.org/0256-307X/35/i=1/a=013701} {\bibfield  {journal}
  {\bibinfo  {journal} {Chinese Physics Letters}\ }\textbf {\bibinfo {volume}
  {35}},\ \bibinfo {pages} {013701} (\bibinfo {year} {2018})}\BibitemShut
  {NoStop}%
\bibitem [{\citenamefont {Liu}\ \emph {et~al.}(2017)\citenamefont {Liu},
  \citenamefont {Vafa},\ and\ \citenamefont {Xu}}]{PhysRevB.95.161116}%
  \BibitemOpen
  \bibfield  {author} {\bibinfo {author} {\bibfnamefont {C.}~\bibnamefont
  {Liu}}, \bibinfo {author} {\bibfnamefont {F.}~\bibnamefont {Vafa}}, \ and\
  \bibinfo {author} {\bibfnamefont {C.}~\bibnamefont {Xu}},\ }\href {\doibase
  10.1103/PhysRevB.95.161116} {\bibfield  {journal} {\bibinfo  {journal} {Phys.
  Rev. B}\ }\textbf {\bibinfo {volume} {95}},\ \bibinfo {pages} {161116}
  (\bibinfo {year} {2017})}\BibitemShut {NoStop}%
\bibitem [{\citenamefont {Chang}\ \emph {et~al.}(2017)\citenamefont {Chang},
  \citenamefont {Xu}, \citenamefont {Zhou}, \citenamefont {Huang},
  \citenamefont {Singh}, \citenamefont {Wang}, \citenamefont {Belopolski},
  \citenamefont {Yin}, \citenamefont {Zhang}, \citenamefont {Bansil},
  \citenamefont {Lin},\ and\ \citenamefont {Hasan}}]{PhysRevLett.119.156401}%
  \BibitemOpen
  \bibfield  {author} {\bibinfo {author} {\bibfnamefont {G.}~\bibnamefont
  {Chang}}, \bibinfo {author} {\bibfnamefont {S.-Y.}\ \bibnamefont {Xu}},
  \bibinfo {author} {\bibfnamefont {X.}~\bibnamefont {Zhou}}, \bibinfo {author}
  {\bibfnamefont {S.-M.}\ \bibnamefont {Huang}}, \bibinfo {author}
  {\bibfnamefont {B.}~\bibnamefont {Singh}}, \bibinfo {author} {\bibfnamefont
  {B.}~\bibnamefont {Wang}}, \bibinfo {author} {\bibfnamefont {I.}~\bibnamefont
  {Belopolski}}, \bibinfo {author} {\bibfnamefont {J.}~\bibnamefont {Yin}},
  \bibinfo {author} {\bibfnamefont {S.}~\bibnamefont {Zhang}}, \bibinfo
  {author} {\bibfnamefont {A.}~\bibnamefont {Bansil}}, \bibinfo {author}
  {\bibfnamefont {H.}~\bibnamefont {Lin}}, \ and\ \bibinfo {author}
  {\bibfnamefont {M.~Z.}\ \bibnamefont {Hasan}},\ }\href {\doibase
  10.1103/PhysRevLett.119.156401} {\bibfield  {journal} {\bibinfo  {journal}
  {Phys. Rev. Lett.}\ }\textbf {\bibinfo {volume} {119}},\ \bibinfo {pages}
  {156401} (\bibinfo {year} {2017})}\BibitemShut {NoStop}%
\bibitem [{\citenamefont {Volovik}\ and\ \citenamefont
  {P.~Mineev}(1977)}]{Volovik1977}%
  \BibitemOpen
  \bibfield  {author} {\bibinfo {author} {\bibfnamefont {G.}~\bibnamefont
  {Volovik}}\ and\ \bibinfo {author} {\bibfnamefont {V.}~\bibnamefont
  {P.~Mineev}},\ }\href@noop {} {\bibfield  {journal} {\bibinfo  {journal}
  {Journal of Experimental and Theoretical Physics - J EXP THEOR PHYS}\
  }\textbf {\bibinfo {volume} {46}},\ \bibinfo {pages} {401} (\bibinfo {year}
  {1977})}\BibitemShut {NoStop}%
\bibitem [{\citenamefont {Bednik}(2019)}]{MyArticle}%
  \BibitemOpen
  \bibfield  {author} {\bibinfo {author} {\bibfnamefont {G.}~\bibnamefont
  {Bednik}},\ }\href {\doibase 10.1103/PhysRevB.100.024420} {\bibfield
  {journal} {\bibinfo  {journal} {Phys. Rev. B}\ }\textbf {\bibinfo {volume}
  {100}},\ \bibinfo {pages} {024420} (\bibinfo {year} {2019})}\BibitemShut
  {NoStop}%
\bibitem [{\citenamefont {Gingras}\ and\ \citenamefont
  {McClarty}(2014)}]{0034-4885-77-5-056501}%
  \BibitemOpen
  \bibfield  {author} {\bibinfo {author} {\bibfnamefont {M.~J.~P.}\
  \bibnamefont {Gingras}}\ and\ \bibinfo {author} {\bibfnamefont {P.~A.}\
  \bibnamefont {McClarty}},\ }\href
  {http://stacks.iop.org/0034-4885/77/i=5/a=056501} {\bibfield  {journal}
  {\bibinfo  {journal} {Reports on Progress in Physics}\ }\textbf {\bibinfo
  {volume} {77}},\ \bibinfo {pages} {056501} (\bibinfo {year}
  {2014})}\BibitemShut {NoStop}%
\bibitem [{\citenamefont {Ferngundez-Pacheco}\ \emph
  {et~al.}(2017)\citenamefont {Ferngundez-Pacheco}, \citenamefont {Streubel},
  \citenamefont {Fruchart}, \citenamefont {Hertel}, \citenamefont {Fischer},\
  and\ \citenamefont {Cowburn}}]{FernGUndez-Pacheco2017}%
  \BibitemOpen
  \bibfield  {author} {\bibinfo {author} {\bibfnamefont {A.}~\bibnamefont
  {Ferngundez-Pacheco}}, \bibinfo {author} {\bibfnamefont {R.}~\bibnamefont
  {Streubel}}, \bibinfo {author} {\bibfnamefont {O.}~\bibnamefont {Fruchart}},
  \bibinfo {author} {\bibfnamefont {R.}~\bibnamefont {Hertel}}, \bibinfo
  {author} {\bibfnamefont {P.}~\bibnamefont {Fischer}}, \ and\ \bibinfo
  {author} {\bibfnamefont {R.~P.}\ \bibnamefont {Cowburn}},\ }\href
  {https://doi.org/10.1038/ncomms15756} {\bibfield  {journal} {\bibinfo
  {journal} {Nature Communications}\ }\textbf {\bibinfo {volume} {8}},\
  \bibinfo {pages} {15756 EP } (\bibinfo {year} {2017})},\ \bibinfo {note}
  {review Article}\BibitemShut {NoStop}%
\end{thebibliography}%

\end{document}